\documentclass{PoS}

\title{The Anomaly Inflow of the domain-wall fermion in odd dimension}

\ShortTitle{Anomaly Inflow in odd dimension}

\author{Hidenori Fukaya\\
        Department of Physics, Osaka University, Toyonaka, Osaka 560-0043, Japan\\
        E-mail: \email{hfukaya@het.phys.sci.osaka-u.ac.jp}}
\author{Naoki Kawai\\
        Department of Physics, Osaka University, Toyonaka, Osaka 560-0043, Japan\\
        E-mail: \email{nkawai@het.phys.sci.osaka-u.ac.jp}}
\author{\speaker{Yoshiyuki Matsuki}\\
        Department of Physics, Osaka University, Toyonaka, Osaka 560-0043, Japan\\
        E-mail: \email{ymatsuki@het.phys.sci.osaka-u.ac.jp}}
\author{Makito Mori\\
        Department of Physics, Osaka University, Toyonaka, Osaka 560-0043, Japan\\
        E-mail: \email{m-mori@het.phys.sci.osaka-u.ac.jp}}
\author{Tetsuya Onogi\\
        Department of Physics, Osaka University, Toyonaka, Osaka 560-0043, Japan\\
        E-mail: \email{onogi@het.phys.sci.osaka-u.ac.jp}}
\author{Satoshi Yamaguchi\\
        Department of Physics, Osaka University, Toyonaka, Osaka 560-0043, Japan\\
        E-mail: \email{yamaguch@het.phys.sci.osaka-u.ac.jp}}

\abstract{In 1985, Callan and Harvey showed a view of gauge anomaly as
a missing current  into an extra-dimension, and the total contribution,
including the Chern-Simons current in the bulk, is conserved.
However in their computation, the edge and bulk contributions are separately
evaluated and their cross correlations, which should be relevant
at boundary, are simply ignored. This issue has been solved in many approaches. In this work, we revisit this issue
with a complete set of eigenstates of free domain-wall Hamiltonian and give the systematic evaluation, easy to take in the higher mass correction and extend to the higher dimension.}

\FullConference{37th International Symposium on Lattice Field Theory - Lattice2019\\
		16-22 June 2019\\
		Wuhan, China}
\begin{document}

\section{Introduction}
It is well-known that the existence of a gauge anomaly leads to an inconsistency of the theory. Therefore, to construct a physical theory, we should pursue gauge anomaly-free theories. Standard Model is an example of anomaly-free theory. It includes some leptons and quarks, whose anomalies cancel with each other. 

The anomaly inflow mechanism proposed in \cite{Callan:1984sa} gives an alternative view to the anomaly cancellation. It views a gauge anomaly as a missing current into the extra-dimension, and there also exist the contribution from the bulk Chern-Simons current, and consequently the total system becomes anomaly-free. To realize this situation, the domain-wall fermion \cite{Kaplan:1992bt, Shamir:1993zy} was used and it produced the edge-localized mode and bulk Chern-Simons action. The anomaly inflow now covers a wide area of physics such as chiral gauge theory on a lattice \cite{Narayanan:1993ss, Jansen:1992yj}, topological matters \cite{Witten:2015aoa, Witten:2015aba} and index theorem \cite{Fukaya:2017tsq, Fukaya:2019qlf, Fukaya:2019myi}. 

In the original work by Callan-Harvey, they treated the anomaly from the bulk and from the edge modes separately. However, what happens near the boundary was not clear. In fact the gauge anomaly should be cancelled even at a microscopic level. To understand the anomaly cancellation at the microscopic level, several studies of the exact effective action were carried out \cite{Jansen:1994ym, Aoki:1993rg, Chandrasekharan:1993ag, Aoki:1996zz, Kawano:1994rg, Hamada:2017tny, Hamada:2018jko} in the infinite bulk mass limit. In this report, we readdress this issue in order to understand inflow mechanics further. In our study, we use the technique based on Ref. \cite{Fukaya:2017tsq, Fukaya:2019qlf, Fukaya:2019myi} computed both the real and imaginary part of the effective action. Our study reveals the cancellation mechanism for both the real and imaginary part. Although, we also take the infinite bulk mass limit, our method allows a systematic treatment of the $1/M$ expansion, although our computation below is limited to its leading order. The application to the higher dimension is also straightforward.

%The anomaly inflow mechanism proposed in \cite{Callan:1984sa} is also one of the anomaly cancellation mechanism. It views a gauge anomaly as a missing current into the extra-dimension, includes the contribution of the bulk Chern-Simons current, and consequently the total system becomes anomaly-free. To realize this situation, the domain-wall fermion \cite{Kaplan:1992bt, Shamir:1993zy} was used and it produced the edge-localized mode and bulk Chern-Simons action. This concept as anomaly cancellation covers a lot of ground, chiral gauge theory on a lattice\cite{Aoki:1996zz},topological matters\cite{Witten:2015aba} and index theorem\cite{Fukaya:2017tsq}. However it is known that this anomaly inflow mechanism has a significant issue -the edge and bulk contributions are separately evaluated and their cross correlations, which should be relevant at boundary, are simply ignored-. A gauge anomaly should be cancelled at a microscopic level. There are some previous works for this issue\cite{Chandrasekharan:1993ag, Kawano:1994rg, Hamada:2017tny, Hamada:2018jko}. But we readdress this issue to evaluate systematically the higher order terms of fermion mass and extend the higher dimension case.

\section{Set-up and Method}
Let us consider the Euclidean 3-dim domain-wall fermion action
\begin{eqnarray}
S=\int dx^3 \overline{\psi}(D\hspace{-6.5pt}/ +M\epsilon(x_{3}))\psi + \sum_{n=1}c_{n}\overline{\phi}_{n}(D\hspace{-6.5pt}/ +M_{n})\phi_{n},
\end{eqnarray}
where $\epsilon(x_{3})$ is the sign function for $x_{3}$-direction and $\phi_{n}$ are the Pauli-Villars regulator field. For simplicity, we consider the spacial variation of back ground $U(1)$ gauge field is small enough compared to the fermion mass scale in the following discussion.

To regulate the loop contributions, the coefficient $c_{n}$ must satisfy the conditions
\begin{eqnarray}
&&\sum_{n=0}c_{n}=0,\ \sum_{n=0}c_{n}|M_{n}|=0,\ \sum_{n=0}c_{n}\frac{M_{n}}{|M_{n}|}=0,
\end{eqnarray}
where $n=0$ corresponds to that of the physical fermion field, that is, $c_{0}=0$ and $M_{0}=M\epsilon(x_{3})$.

\begin{figure}[h]
\begin{center}
\begin{tabular}{c}
\begin{minipage}{0.33\hsize}
\begin{center}
  \includegraphics[clip,width=6.0cm]{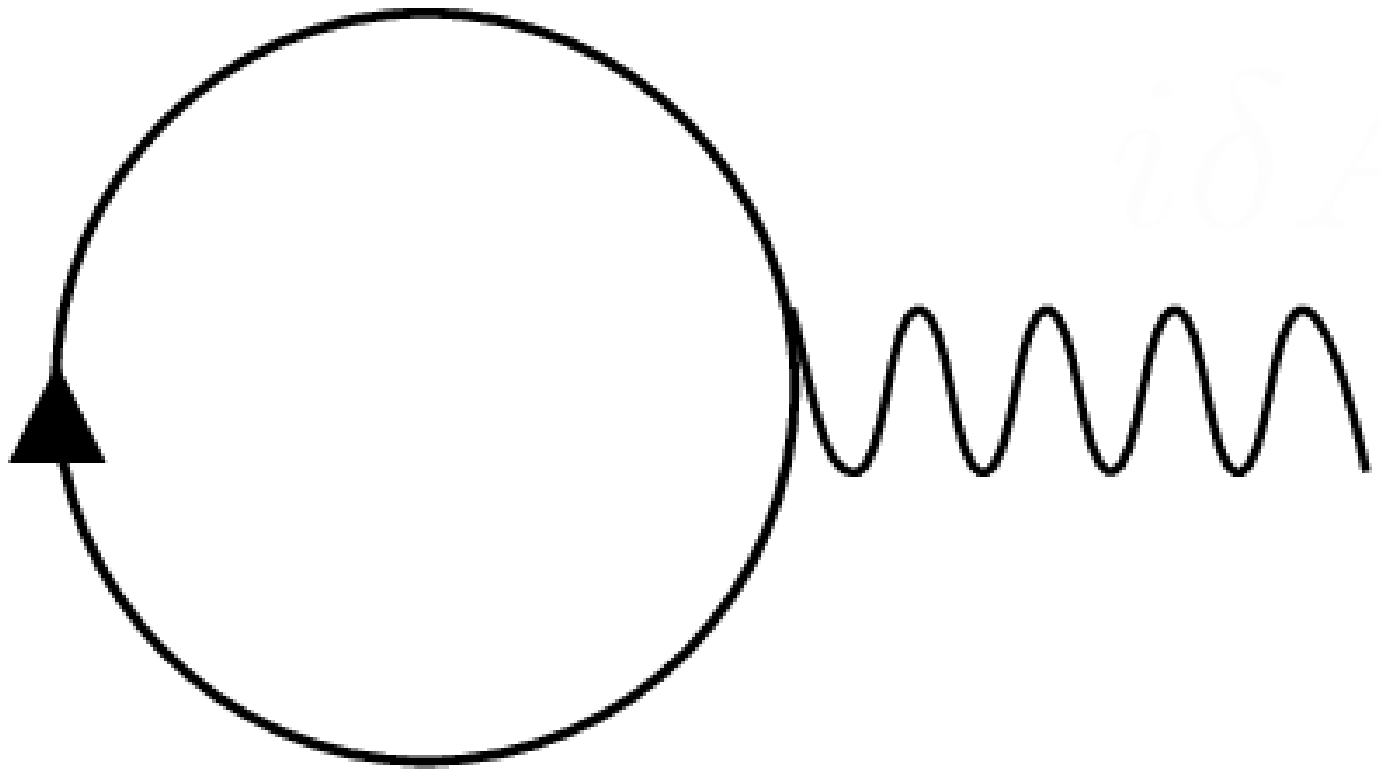}
  \caption{one point function}
  \label{fig:one}
  \end{center}
  \end{minipage}

\begin{minipage}{0.33\hsize}
\begin{center}
  \includegraphics[clip,width=6.0cm]{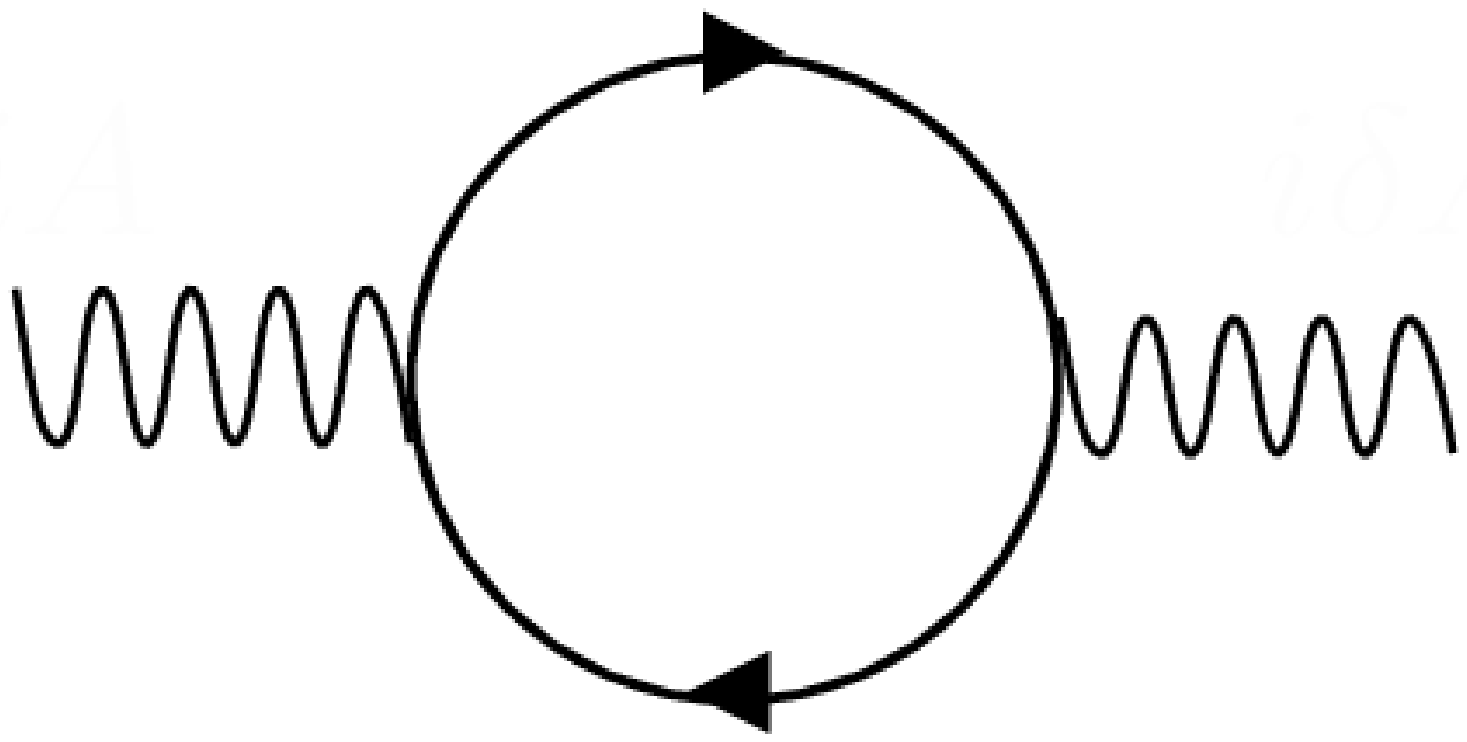}
  \caption{two point function}
  \label{fig:two}
  \end{center}
  \end{minipage}
  \end{tabular}
  \end{center}
\end{figure}

Our goal is to confirm the gauge anomaly cancellation at the microscopic level, that is, the local current conservation. For this purpose, we evaluate the variation of the effective action
\begin{eqnarray}
\delta S_{\mathrm{eff}}&=&\mathrm{Tr}\left(\frac{i\delta A\hspace{-6pt}/ }{D\hspace{-6.5pt}/ +M\epsilon(x_{3})}\right)+PV.\\
\label{eq:1}
&=&\mathrm{Tr}[S_{0}i\delta A\hspace{-6pt}/ -S_{0}i A\hspace{-6pt}/S_{0}i\delta A\hspace{-6pt}/ +\cdots].
\end{eqnarray}
In the last line, we expand in powers of gauge filed. $S_{0}$ is the free domain-wall propagator
\begin{eqnarray}
S_{0}=\frac{1}{\partial \hspace{-6.5pt}/ +M\epsilon(x_{3})}.
\end{eqnarray}
Moreover, we decompose this propagator by using the a complete set of eigenstates of free domain-wall Hamiltonian to see explicitly the cross correlation between the bulk modes and edge modes. Let us consider a free domain-wall Hamiltonian
\begin{eqnarray}
H=-\partial^2+M^2+2\gamma^{3}\delta(x_{3}) M. 
\end{eqnarray}
The eigenmodes for this hamiltonian are
\begin{eqnarray}
&&\phi^{Bulk,e}_{\eta}(x)=\frac{1}{\sqrt{4\pi(k_{3}^2+M^2)}}((k_{3}-\eta iM)e^{-ik_{3}|x_{3}|}+(k_{3}+\eta iM)e^{ik_{3}|x_{3}|})e^{i\vec{k}\cdot\vec{x}}u_{\eta},\\
&&\phi^{Bulk,o}_{\eta}(x)=\frac{1}{\sqrt{4\pi}}(e^{-ik_{3}x_{3}}-e^{ik_{3}x_{3}})e^{i\vec{k}\cdot\vec{x}}u_{\eta},\\
&&\phi^{Edge,e}_{+}(x)=\sqrt{M}e^{-M|x_{3}|}e^{i\vec{k}\cdot\vec{x}}u_{+},\\
&&u_{+}=\left(\begin{array}{c}
1\\
0
\end{array}
\right), u_{-}=\left(\begin{array}{c}
0\\
1
\end{array}
\right),
\end{eqnarray}
where $\eta=\pm$ represents the chirality (the eigenvalues of $\gamma^{3}$), $e/o$ means the even/odd function for $x_{3}$, and $\vec{k}=(k_{1},k_{2})$.
We can see a complete set is composed of the bulk modes and the edge localized modes. Therefore, the propagator can be decomposed in the preferred form
\begin{eqnarray}
S_{0}(x,y)&=&\int_{\vec{k}} \frac{1}{2E_{k}}\left[\gamma^{3}E_{k}\epsilon(x_{3}-y_{3})-i\vec{k\hspace{-5pt}/}+M\epsilon(y_{3})\right]e^{-E_{k}|x_{3}-y_{3}|+i\vec{k}\cdot(\vec{x}-\vec{y})}\nonumber \\
&&-M\int_{\vec{k}} \frac{1}{2E_{k}}\left[\epsilon(y_{3})+(\gamma^{3}E_{k}+M)\frac{i\vec{k\hspace{-5pt}/}}{\vec{k}^2}\right]e^{-E_{k}(|x_{3}|+|y_{3}|)+i\vec{k}\cdot(\vec{x}-\vec{y})},\\
&\equiv& S^{B}_{0}+S^{E}_{0},
\end{eqnarray}
where $\int_{\vec{k}}=\int d^2\vec{k}/(2\pi)^2$, $E_{k}=\sqrt{\vec{k}^2+M^2}$ and $S^{B}_{0},S^{E}_{0}$ denote the propagator originated from the bulk-modes and the edge-modes respectively. By using above decomposition, we will evaluate the one-point and two-point functions in (\ref{eq:1}) (Figures. \ref{fig:one},\ref{fig:two}), and see the local anomaly cancellation between the bulk modes and the edge modes. We will explain the results in the succeeding section.
\section{Results}
From now, we show the results of one- and two-point functions since the anomaly does not come from the after three-point function in our set-up.
\subsection{one-point function}
\begin{eqnarray}
\mathrm{Tr}\left[S_{0}i\delta A\hspace{-6pt}/ \right]=
\mathrm{tr}\int d^3 x \int_{\vec{k}}\frac{1}{2E_{k}}\left[i\vec{k\hspace{-5pt}/}+M\epsilon(x_{3})-M\left(\epsilon(x_{3})+(\gamma^{3}E_{k}+M)\frac{i\vec{k}}{\vec{k}^2}\right)e^{-2E_{k}|x_{3}|}\right]i\delta A\hspace{-6pt}/(x).
\end{eqnarray}
This is the result of the one-point function. The first and second terms in the square brackets come from the bulk modes and the remaining terms do from the edge modes. This becomes $0$ due to the odd functions for $\vec{k}$ and the property of trace. So the one-point function does not contribute to the anomaly.
\subsection{two-point function}
\begin{figure}[h]
\begin{center}
  \includegraphics[clip,width=7.0cm]{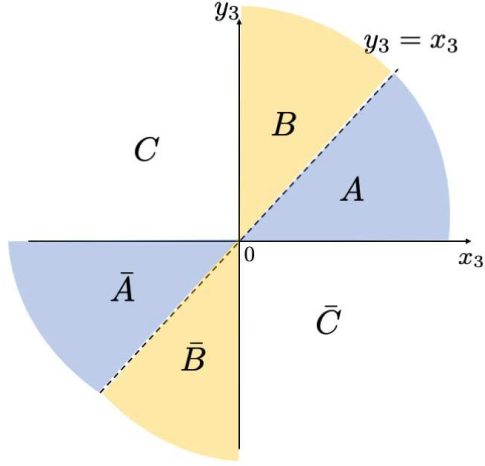}
  \caption{The division of the $x_{3}$-$y_{3}$ space. The regions $A$ and $\bar{A}$ ($B$ and $\bar{B}$, $C$ and $\bar{C}$) are almost same calculation.}
  \label{fig:oma}
  \end{center}
\end{figure}

\begin{eqnarray}
\mathrm{Tr}\left[S_{0}iA\hspace{-6pt}/ S_{0}i\delta A\hspace{-6pt}/ \right]=\sum_{i,j=B,E}\mathrm{Tr}\left[S_{0}^{i}iA\hspace{-6pt}/ S^{j}_{0}i\delta A\hspace{-6pt}/ \right].
\end{eqnarray}
As we mentioned, the propagator can be decomposed into the contribution of the bulk and the edge. First, we explain the basic idea of our evaluation. In calculating the two-point function, we encounter some expressions including the sign function $\epsilon(x_{3})$, e.g. 
\begin{eqnarray}
\label{eq:2}
\int dx_{3}dy_{3} e^{-(E_{k}+E_{p})|x_{3}-y_{3}|}\epsilon(x_{3})\epsilon(y_{3})f(x_{3},y_{3}),
\end{eqnarray}
where $f(x_{3},y_{3})$ is the analytic function which does not include sign functions. We can evaluate this by dividing the $x_{3}$-$y_{3}$ space into six areas (four areas divided by $y_{3}=x_{3}$ in 1st and 3rd quadrants, 2nd and 4th quadrants), like Figure \ref{fig:oma}. And then, we can carry out the evaluation of this expression as follows. 
\begin{eqnarray}
(\ref{eq:2})&=&\int_{0}^{\infty}dx_{3}\sum_{n}\frac{1}{n !}(f^{(n,0)}(x_{3},x_{3})+f^{(0,n)}(x_{3},x_{3}))\Gamma (n+1)\frac{1}{(E_{k}+E_{p})^{n+1}}\nonumber\\
&&+\int_{-\infty}^{0}dx_{3}\sum_{n}\frac{1}{n !}(-1)^{n}(f^{(n,0)}(x_{3},x_{3})+f^{(0,n)}(x_{3},x_{3}))\Gamma (n+1)\frac{1}{(E_{k}+E_{p})^{n+1}}\nonumber\\
&&-\sum_{n,m}\frac{1}{n !m!}(-1)^{n}(f^{(n,m)}(0,0)+f^{(m,n)}(0,0))\Gamma (n+1)\Gamma (m+1)\frac{1}{(E_{k}+E_{p})^{n+m+1}},
\end{eqnarray}
where $n,m$ runs from $0$ to $\infty$ and $f^{(n,m)}$ means the $n$-th derivative with respect to the first argument and $m$-th derivative with respect to the second argument. Since $E_{k}$ includes the fermion mass $M$, our method allows to evaluate systematically the higher order terms in 1/M expansion by taking in the contribution of the higher $n,m$.

Other expressions including the sign functions can be evaluated in the same manner. Furthermore, this method can be extended to the higher dimension case because this method depends on only the property of sign function. For example, if you want to consider the 5 dimension case, one more sign function is added and the number of division of space increase. But, the procedure is same as 3 dimension case. This is why our approach is systematic.

 Let us evaluate the contribution from only the bulk modes.
\subsubsection*{Bulk-Bulk contribution}
\begin{eqnarray}
\mathrm{Tr}&&\left[S_{0}^{B}iA\hspace{-6pt}/ S^{B}_{0}i\delta A\hspace{-6pt}/ \right]\nonumber\\
&&=-\frac{i}{4}\int d^{3}x F_{\mu\nu}\delta A_{\rho}\theta(x_{3})\epsilon^{\mu\nu\rho}-\frac{i}{4}\int dx_{1}dx_{2}A_{\mu}\delta A_{\nu}\epsilon^{\mu\nu 3}\bigr|_{x_{3}=0}-\frac{1}{8\pi}\int dx_{1}dx_{2}A_{\mu}\delta A^{\mu}\bigr|_{x_{3}=0}. \nonumber \\
\end{eqnarray}
This is the result in the leading order. We note that this has not only a usual imaginary anomalous term but also a mass term breaking the gauge invariance localized at $x_3=0$.
Next, we show the result of the cross correlation between the bulk modes and the edge modes.

\subsubsection*{Bulk-Edge cross correlation contribution}
The result of the cross correlation between the bulk modes and the edge modes can be obtained in almost same procedure as the previous calculation, so we show the result in leading order and give some comments.
\begin{eqnarray}
\mathrm{Tr}\left[S_{0}^{B}iA\hspace{-6pt}/ S^{E}_{0}i\delta A\hspace{-6pt}/ \right]+\mathrm{Tr}\left[S_{0}^{E}iA\hspace{-6pt}/ S^{B}_{0}i\delta A\hspace{-6pt}/ \right]=\frac{1}{4\pi}\int dx_{1}dx_{2} A_{3}\delta A_{3}\bigr|_{x_{3}=0}-\frac{1}{8\pi}\int dx_{1}dx_{2} A_{i}\delta A^{i}\bigr|_{x_{3}=0},
\end{eqnarray}
where $i$ runs 1,2. This expression has also a usual imaginary anomalous term and a mass term, which is known in \cite{Kawano:1994rg}. Therefore, the contribution from the cross correlation between the bulk modes and the edge modes ignored in \cite{Callan:1984sa} is important.
Finally, we show the result of the contribution of only the edge modes. 

\subsubsection*{Edge-Edge contribution}
\begin{eqnarray}
\mathrm{Tr}&&\left[S_{0}^{E}iA\hspace{-6pt}/ S^{E}_{0}i\delta A\hspace{-6pt}/ \right]\nonumber\\
&&=-\frac{1}{4\pi}\int dx_{1}dx_{2} A_{i}\delta A^{i}\bigr|_{x_{3}=0}-\frac{1}{8\pi}\int dx_{1}dx_{2} A_{3}\delta A^{3}\bigr|_{x_{3}=0}+\frac{1}{2\pi}\int dx_{1}dx_{2} \delta A_{j}\frac{\partial^{i}\partial^{j}}{\partial^2}A_{i}\bigr|_{x_{3}=0}\nonumber\\
&&+\frac{i}{4\pi}\int dx_{1}dx_{2} \epsilon^{jk3}\delta A_{k}\frac{\partial^{i}\partial^{j}}{\partial^2}A_{i}\bigr|_{x_{3}=0}+\frac{i}{4\pi}\int dx_{1}dx_{2} \epsilon^{ik3}\delta A_{j}\frac{\partial^{i}\partial^{j}}{\partial^2}A_{k}\bigr|_{x_{3}=0},
\end{eqnarray}
where $i,j,k$ runs 1,2. The edge localized mode is the gapless mode so the non-local term appears. Moreover, we see that the edge contribution also has a gauge-variance mass term. 
\subsubsection*{Total contribution}
Summing up these results, we obtain the following expression
\begin{eqnarray}
\delta S_{\mathrm{eff}}&&=-\frac{i}{4\pi}\int d^3 x F_{\mu\nu}\delta A_{\rho}\epsilon^{\mu\nu\rho}\theta(x_{3})-\frac{i}{4\pi}\int dx_{1}dx_{2} A_{\mu}\delta A_{\nu} \epsilon^{\mu\nu 3}\bigr|_{x_{3}=0}\nonumber\\
&&+\frac{1}{2\pi}\int dx_{1}dx_{2} \delta A_{j}\frac{\partial^{i}\partial^{j}-\partial^{2}\delta_{ij}}{\partial^2}A_{i}\bigr|_{x_{3}=0}+\frac{i}{4\pi}\int dx_{1}dx_{2} \epsilon^{jk3}\delta A_{k}\frac{\partial^{i}\partial^{j}}{\partial^2}A_{i}\bigr|_{x_{3}=0}\nonumber\\
&&+\frac{i}{4\pi}\int dx_{1}dx_{2} \epsilon^{ik3}\delta A_{j}\frac{\partial^{i}\partial^{j}}{\partial^2}A_{k}\bigr|_{x_{3}=0}.
\end{eqnarray}
The mass terms are exactly canceled out with each other. Furthermore, we confirm the local current conservation
\begin{eqnarray}
\partial_{\mu}J^{\mu}=-\partial_{\mu}\frac{\delta S_{\mathrm{eff}}}{\delta A_{\mu}}=\frac{i}{4\pi}F_{\mu\nu}\epsilon^{\mu\nu 3}\delta(x_{3})-\frac{i}{4\pi}\partial_{\mu}A_{\nu}\epsilon^{\mu\nu 3}\delta(x_{3})-\frac{i}{4\pi}\partial_{\mu}A_{\nu}\epsilon^{\mu\nu 3}\delta(x_{3})=0.
\end{eqnarray}
Namely, the gauge anomaly is canceled at the microscopic level, which is known in\cite{Chandrasekharan:1993ag, Kawano:1994rg, Hamada:2017tny}.

The effective action is obtained as
\begin{eqnarray}
S_{\mathrm{eff}}=-\frac{i}{8\pi}d^3 x F_{\mu\nu}A_{\rho}\theta(x_{3})\epsilon^{\mu\nu\rho}+\frac{1}{8\pi}\int dx_{1} dx_{2} F^{ij}\frac{1}{\partial^2}F_{ij}-\frac{i}{8\pi}\int dx_{1} dx_{2} \epsilon^{ij3}\partial_{k}A^{k}\frac{1}{\partial^2}F_{ij},
\end{eqnarray}
where the first term is well-known Chern-Simons action, the second term is the gauge-invariance non-local term which does not contribute to the anomaly, and the final non-local term originates from the edge localized modes. The first and last term produce the anomalous term but these terms cancel out.
\section{Conclusion}
We have readdressed the issue of the anomaly inflow mechanism \cite{Callan:1984sa}, in which the edge and bulk contributions are separately evaluated and their cross correlations are simply ignored. The solutions of this issue was already known in \cite{Chandrasekharan:1993ag, Kawano:1994rg, Hamada:2017tny}, but we have tried to evaluate the effective action systematically in order to take the higher mass correction and the higher dimension into account. We have made the formulae for the integrand including the sign functions, which depends on the mathematical property of sign functions and are  systematically applicable even though the number of the sign functions increases. Finally, we have calculated the variation of the effective action by using the formulae, and have confirmed that the anomaly inflow mechanism is exactly correct at the microscopic level in domain-wall fermion set up.
\section*{Acknowledgment}
This work was supported in part by JSPS KAKENHI Grant Number JP15K05054, JP18H01216, JP18H04484, JP18K03620, and JP19J20559.

\end{document}